\documentstyle[epsfig,preprint,prb,aps]{revtex}
\newcommand {\dV}     {{\delta V}}

\newcommand {\UNIV}   {Universit\`a }
\begin{document}
\title{Intermittency and scaling laws for wall bounded turbulence}
\author{
R\@. Benzi
\thanks{AIPA, via Solferino 15, 00185, Roma, on leave of absence from
        Dip. di Fisica, \UNIV di Roma ``Tor Vergata'', Italy.},
G\@. Amati
\thanks{CASPUR, p.le A. Moro 5, 00185 Roma, Italy.},
C\@.M\@. Casciola
\thanks{Dip. Mecc. Aeron., \UNIV di Roma "La Sapienza", 
        via Eudossiana 18, 00184, Roma, Italy.},
F\@. Toschi
\thanks{Dip. di Fisica, \UNIV di Pisa, piazza Torricelli 2,
        56126, Pisa, Italy}
\&
R\@. Piva$^\ddag$.}
\maketitle
\begin{abstract}
Well defined scaling laws clearly appear in wall bounded turbulence, 
even very close to the wall, where a distinct violation of the refined 
Kolmogorov similarity hypothesis (RKSH) occurs together with the
simultaneous persistence of scaling laws. 
A new form of RKSH for the wall region is here proposed in terms of the 
structure functions of order two which, in physical terms, confirms the 
prevailing role of the momentum transfer towards the wall in the near wall 
dynamics.
\end{abstract}

\newpage


\vspace*{1.cm}

The intermittent behavior of velocity increments in the inertial range
of fully developed turbulence has been a subject of renewed interest during 
the years, starting from the objection that Landau raised to Kolmogorov theory 
of 1941 (K41).
Since then, any theory of the inertial range can not avoid considering
the effect of intermittent dissipation of energy on the inertial scales
of motion. Under this respect, the Kolmogorov-Obukhov refined similarity
hypothesis (RKSH), certainly the most credited \cite{FRISH}, leads to 
a probability distribution function of velocity increments characterized by 
the scaling

\begin{eqnarray}
\label{RKSH}
< \dV^p > & \propto & < \epsilon_r^{p/3} > r^{p/3} \, ,
\end{eqnarray}

\noindent where $\epsilon_r^{q}$  denotes the $q^{th}$ moment of the 
dissipation spatially averaged over a volume of characteristic dimension $r$
and the brackets indicate ensemble averaging. 
Taking into account the scaling properties of the dissipation field,

\begin{eqnarray}
\label{E_SC}
< \epsilon_r^{q} > &  \propto & r^{\tau(q)} \, ,
\end{eqnarray}

\noindent equation (\ref{RKSH}) implies that the velocity structure function of
order $p$ is expressed as a power law of the separation with exponent

\begin{eqnarray}
\label{zeta_p}
\zeta_p & = & \tau(p/3) \, + \, p/3 \,.
\end{eqnarray}

\noindent Here, the anomalous correction, $\tau(p/3)$, to the  K41-exponent 
accounts for the intermittency of the velocity increments in the inertial
range of homogeneous and isotropic turbulence. 

A substantial  extension of the range of scales where similarity is observed
has recently been achieved \cite{PhysD} by assuming, as basic quantity, the 
third order structure function instead of the separation $r$, 
\begin{eqnarray}
\label{GESS}
< \dV^p > & \propto & \frac{< \epsilon_r^{p/3} >} {< \epsilon >^{p/3}}
\, < \dV^3 >^{p/3} \,,
\end{eqnarray}

\noindent as suggested by the Kolmogorov equation \cite{FRISH}.
A direct consequence of eq. (\ref{GESS}) is the existence of an extended 
self-similarity (ESS) of the generic structure function of order $p$ in terms
of the third order moment with exponent $\zeta_p$.
Since its introduction, the generalized Kolmogorov similarity 
hypothesis (\ref{GESS}) has appeared as the characteristic feature of 
a  vast number of turbulent systems.

%
%

\begin{figure} [!hb]
\vspace*{1.cm}
   \centerline{
   \epsfig{figure=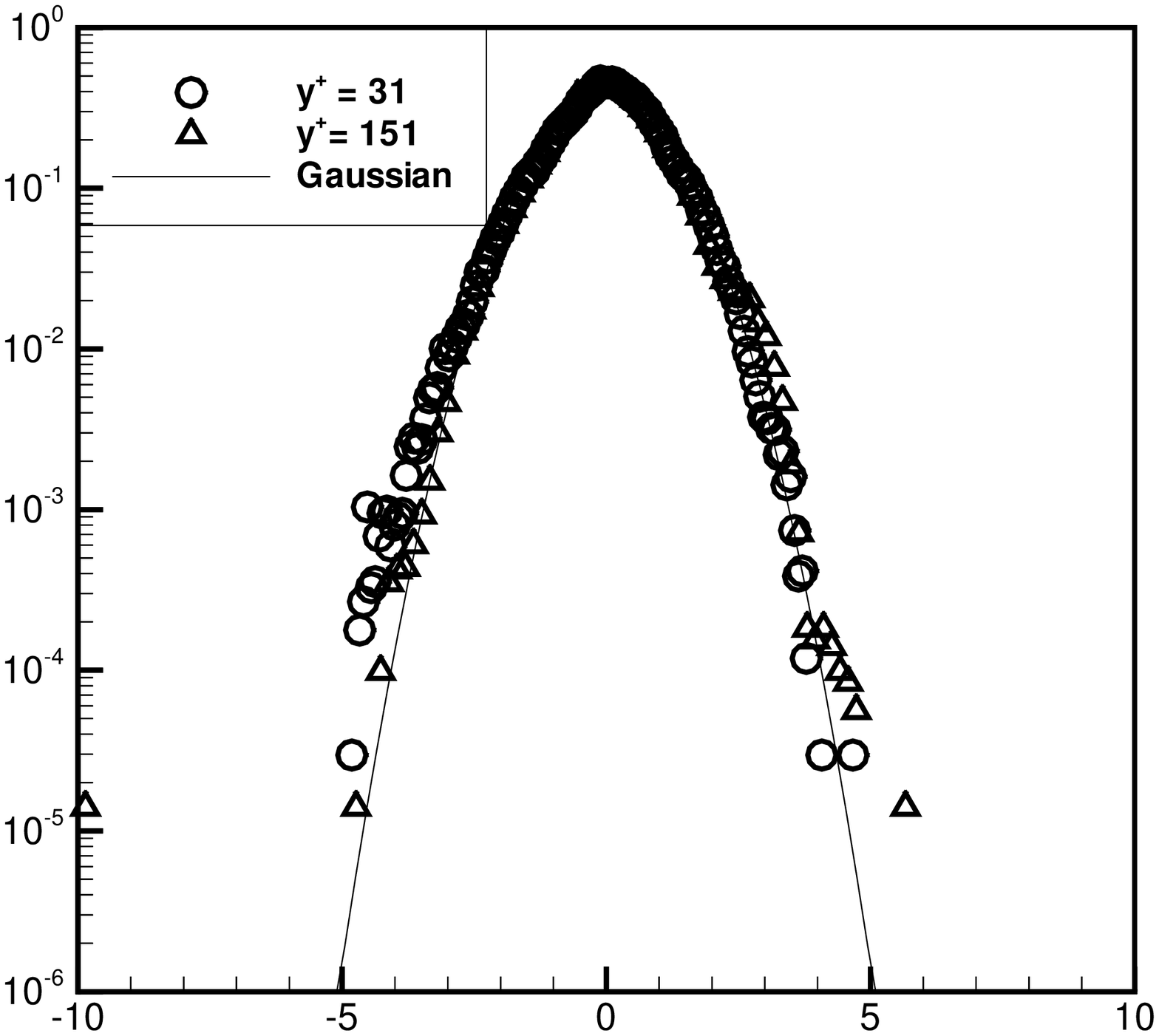,width=4.5cm}
   \epsfig{figure=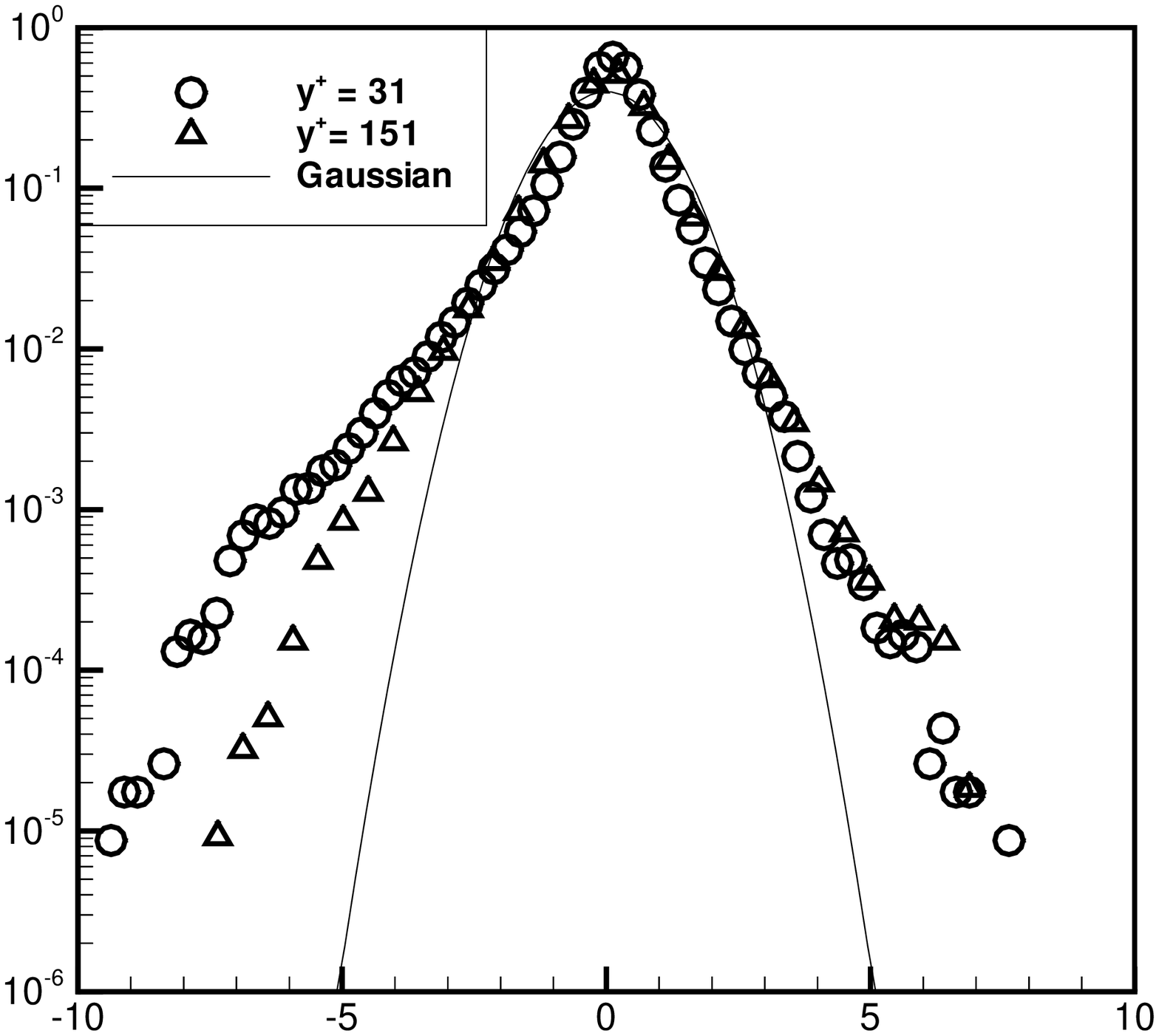, width=4.5cm} }
\vspace*{1.cm}
\caption{Pdf of the velocity increments for different values of the separation 
(left $r^+ = 160$, right $r^+=18$) at two distances from the wall: 
$y^+ = 151$, near the center of the channel,  and $y^+=31$, in the wall region 
of the flow. 
Data from DNS of a turbulent channel flow with $Re_* =160$ 
{\protect \cite{FDR}}. Wall units are used throughout the paper.
\label{F1}}
\end{figure}

In the present letter we intend to discuss the issue of intermittency in wall 
bounded turbulence and its relationship with scaling (ESS) laws, which have 
been observed \cite{FDR},\cite{Nizza} even in regions very close to the wall 
dominated by quite ordered vortical structures \cite{CA}.
As shown in fig. (\ref{F1}), we have evidence that intermittency 
increases moving from the bulk of the fluid towards the wall 
\cite{Nizza}. 
In principle, one may attempt to describe this behavior in the framework of 
RKSH, in its generalized form (\ref{GESS}).
Hence the larger intermittency (smaller $\zeta_p$) would be provided by an 
increase of intermittent fluctuations of $\epsilon_r$ (larger values of 
$| \tau(p) |$). In such conditions, the anomaly of the scaling exponents
would strongly depend on the local flow properties, loosing, thus, any 
trait of universality.

\begin{figure} [!ht]
\vspace*{1.cm}
\epsfysize5.0cm
\epsfxsize8.0cm
\centerline{
 \leavevmode\epsfbox{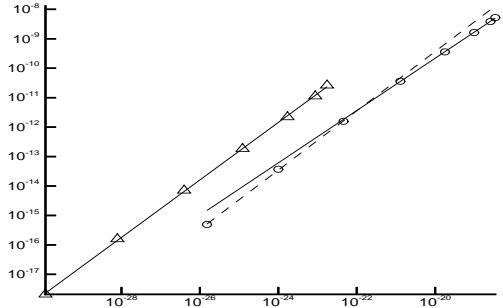} }
\vspace*{1.cm}
\caption{ $<\dV^6>$ vs $<\epsilon^2> \, <\dV^3>^2$ for two different wall 
normal distances.
Bulk region ($y^+ = 151$):
data (triangles) and their fit in the region $r^+ \in [20,\, 320]$ (solid line 
with slope $.99$).
Wall region ($y^+ = 31$): data (circles) and their fits in the two regions 
$r^+ \in [1, \,  20]$  and $r^+ \in [20, \, 320]$, solid line with slope 
$.99$ and dotted line with slope $0.88$, respectively.
\label{F3}}
\end{figure}

To assess the self-consistency of this approach, 
in fig (\ref{F3}) we plot on a logarithmic scale the structure function
of order six versus $<\epsilon^2> \, <\dV^3>^2$.
On the basis of the assumed
validity of (\ref{GESS}), the plot should result in a straight line of slope 
$s=1$, independent of the distance from the wall. This behavior actually 
emerges near the center of the channel  while in the wall region 
a quite clear, though small, violation is manifested.
Specifically, for $y^+ = 31$ two different scaling laws appear. The one,
characterized by slope $s=1$, trivially pertains to the dissipative range. 
The other, with slope $s=.88$, which doesn't satisfy (\ref{GESS}), shows
a first clear example of failure of RKSH.

The previous discussion may suggest a relationship between the increase of 
intermittency, observed in the near wall region, and the simultaneous breaking 
of the RKSH. To this 
regard, it seems interesting to investigate the possible existence of a new 
form of RKSH valid in the near wall region.
In fact RKSH, somehow suggested by the well known ``$4/5$'' Kolmogorov equation 
(see Frish \cite{FRISH}), tells us, in physical terms,  that the 
``energy flux'' in the inertial range, represented by the term $(\delta V_r)^3$, 
fluctuates with a probability distribution which is the same of $\epsilon_r$. 
However, in the case of strong shear, we should expect that a new term,
proportional to $\partial_z<U> (\delta V_r)^2$, enters the
estimate of the energy flux at scale $r$. 
Such a new term, indeed, appears in the analysis performed for homogeneous 
shear flows (see for instance Hinze \cite{HINZE}). 
If this term becomes dominant, as it may occur for a very large shear, 
one is led to assume that the fluctuations of the energy
flux in the inertial range are proportional to $(\delta V_r)^2$, {\sl i.e.}
$\epsilon_r \propto A(r) (\delta V_r)^2$, with $A(r)$ a non fluctuating 
function of $r$. 
Hence, we may expect that a new form of the RKSH should hold which, in its 
generalized form, reads as
\begin{eqnarray}
\label{new}
< \dV^p > & \propto & \frac{< \epsilon_r^{p/2} >} {< \epsilon >^{p/2}}
\, < \dV^2 >^{p/2} \,.
\end{eqnarray}
\noindent The above expression of the new RKSH is given in terms of the
structure function of order two, without explicit reference to the separation 
$r$, in the same way as the generalized RKSH (\ref{GESS}).
In the spirit of the extended self similarity, we assume the new form of RKSH 
to be valid also in the region very close to the wall, where the shear is 
certainly prevailing.

\begin{figure} [!hb]
\vspace*{1.cm}
\begin{center}
\epsfysize5.0cm
\epsfxsize8.0cm
\leavevmode\epsfbox{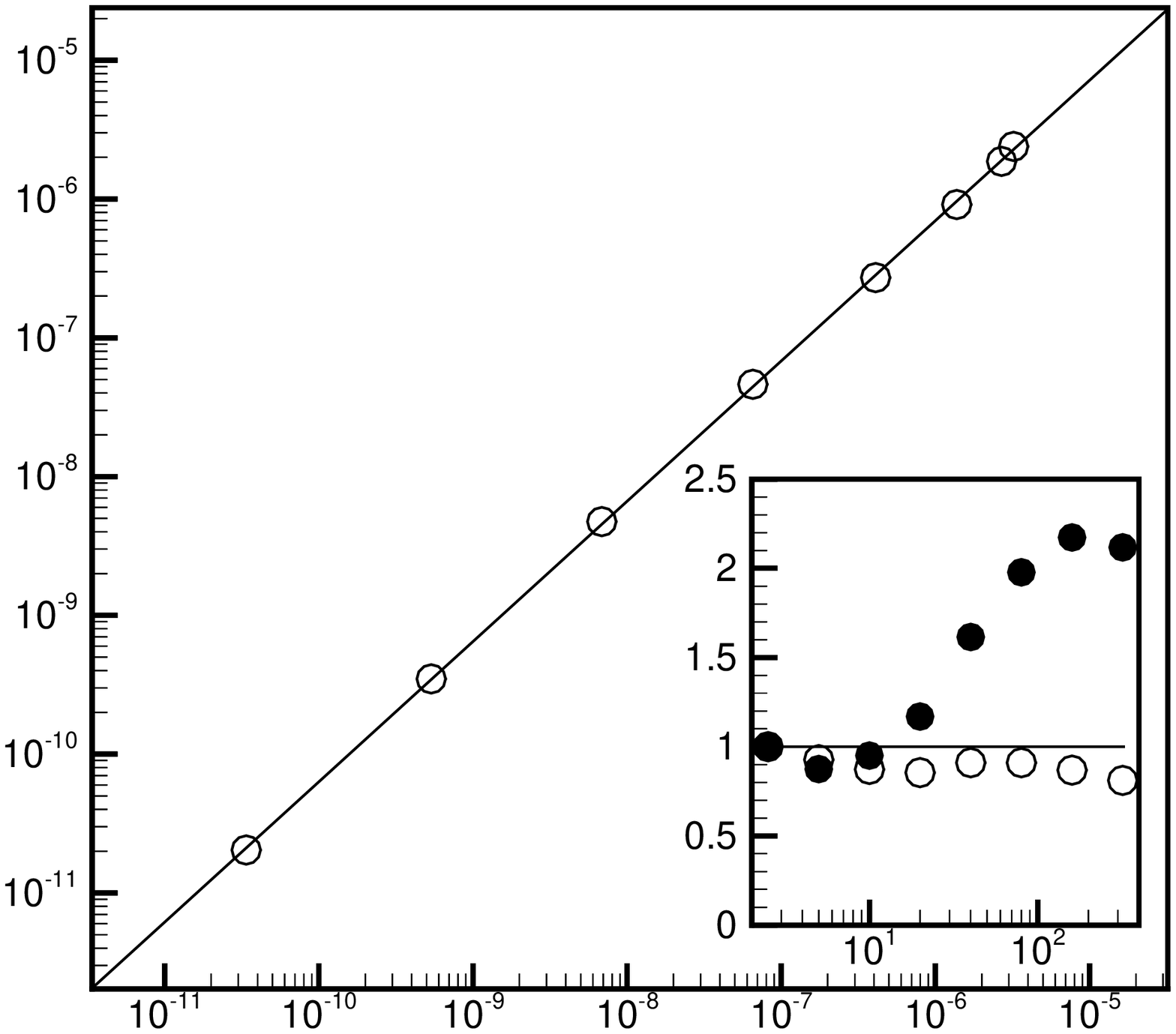}
\vspace*{1.cm}
\caption{Check of consistency for eq. (\ref{new}) at $y^+= 31$:
 $<\dV^4>$ vs $<\epsilon^2> \, <\dV^2>^2$.
The solid line (slope $ 1.01$) gives the fit in the whole range.
In the insert: 
open circles, compensated plot for eq. (\ref{new}), 
$<\dV^4> \, / <\epsilon^2> \, <\dV^2>^2 \,$ vs $\, r^+$;
filled circles, corresponding  plot for eq. (\ref{GESS}), 
$ <\dV^6>\, / <\epsilon^2> \, <\dV^3>^2 \,$ vs $\, r^+$.
\label{F5.a}}
\end{center}
\end{figure}

In order to verify this set of assumptions, we show in fig. (\ref{F5.a}) 
a log-log plot of equation (\ref{new}) for $p=4$ at $y^+ = 31$.
In the insert, we show for the same plane the compensated plot of both 
(\ref{new}) for $p=4$ and (\ref{GESS}) for $p=6$. 
It follows a quite clear agreement of eq.  (\ref{new})  with the 
numerical data.  In principle, the function $A(r)$ might be evaluated
theoretically starting from the Kolmogorov
equation for anisotropic shear flow (e.g. see \cite{PRO}).

\begin{figure} [!hb]
\vspace*{1.cm}
\begin{center}
\epsfysize5.0cm
\epsfxsize8.0cm
\leavevmode\epsfbox{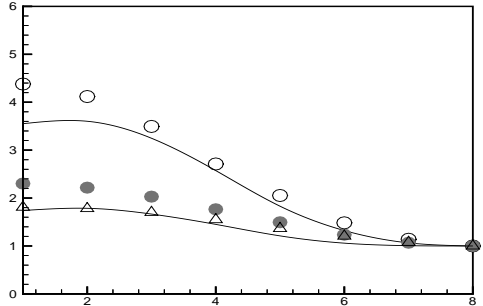}
\vspace*{1.cm}
\caption{Flatness, $ F = < \dV^4 >/<\dV^2>^2 $) vs $\log_2(r^+/Dx^+)$,
$Dx^+=2.5$, at $y^+ = 151$ (open triangles) and $y^+=31$ (open circles), as 
evaluated by eqs. (\ref{F_w}), using $F_b$ and $F_w$, respectively. 
For comparison: filled circles, $F_b$ applied at $y^+=31$. 
Correspondingly, the solid lines give the flatness as evaluated directly in 
terms of velocity.
\label{F6}}
\end{center}
\end{figure}

\begin{figure} [!ht]
\vspace*{1.cm}
 \begin{center}
 \epsfysize5.0cm
 \epsfxsize8.0cm
 \leavevmode\epsfbox{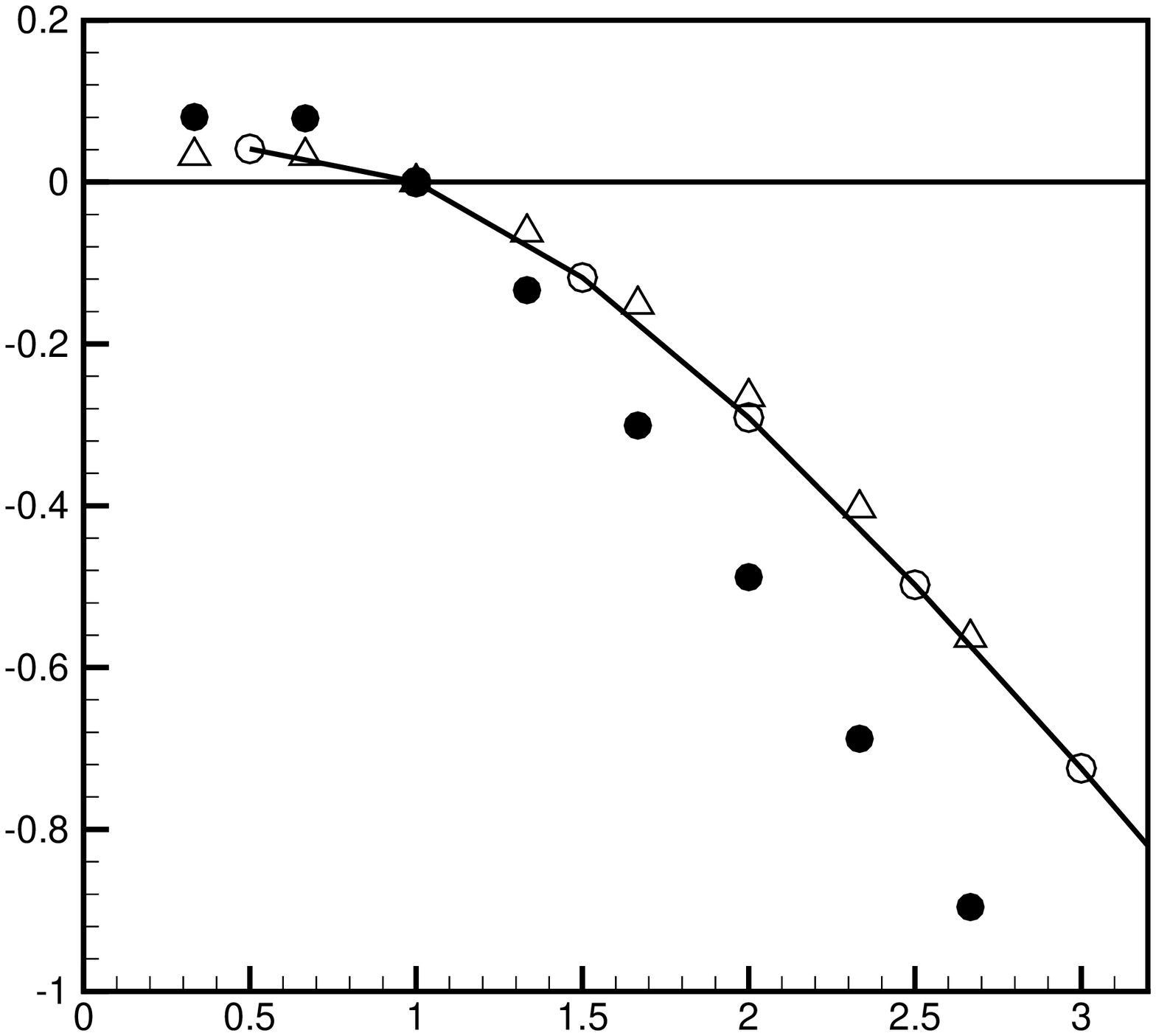}
\vspace*{1.cm}
 \caption{The anomalous correction $\hat \tau(q)$ as computed by the new
 scaling law for the wall region ($y^+= 31 $), eq. (\ref{new}),  
 (open circles) compared with that issuing from RKSH at both $y^+ = 151 $ 
 (triangles) and $y^+ = 31 $  (filled circles). 
 \label{F5.b}}
 \end{center}
 \end{figure}

The increased intermittency of the velocity fluctuations near the wall
may be estimated by considering how the flatness $F(r)$ grows with 
$r \rightarrow 0$, with
\begin{eqnarray}
\label{F_a} 
F(r) & = &  \frac{<\dV^4(r)>}{<\dV^2(r)>^2}  \, .
\end{eqnarray}
\noindent 
By combining the definition (\ref{F_a}) with (\ref{GESS}) and (\ref{new}) 
we obtain the following expressions in terms of $\epsilon_r$,

\begin{eqnarray}
\label{F_w}
F_b  \, = \, \frac{<\epsilon_r^{4/3}>}{<\epsilon_r^{2/3}>^2} & \qquad \qquad & 
F_w  \, = \, \frac{<\epsilon_r^{4/2}>}{<\epsilon_r^{2/2}>^2}  \,,
\end{eqnarray}

\noindent  which are suitable for the bulk and near the wall region, 
respectively.
As we see from fig. (\ref{F6}), both $F_b$ and $F_w$ diverge for 
$r \rightarrow 0$, indicating intermittent behavior in both cases,
if we exclude the smallest separations falling into the dissipative range.
Clearly $F_w$ diverges faster than $F_b$.
This result is consistent with the corresponding analysis performed 
directly in terms of structure functions of velocity by means of 
eq. (\ref{F_a}) and provides a further evidence of the validity of (\ref{new})
near the wall.
In fact, the application of $F_b$ near the wall doesn't 
catch the increase of intermittency of the velocity fluctuations (see fig. 
(\ref{F6})).
On the other hand, the differences in the statistical properties of the 
dissipation between the bulk and the near wall region are too small to account 
for the increase of intermittency of the velocity increments near the wall. 
This is indirectly confirmed by the observed direct scaling (ESS) of the 
structure functions with $< \dV^3 >$, which implies, starting from
eq. (\ref{new}), 

\begin{eqnarray}
\label{new_tau_p}
\hat \tau(p/2) & = & 
\hat \zeta_p  - \frac{p}{2} \, \hat \zeta_2 \,,
\end{eqnarray}

\noindent where a hat has been introduced here to denote the scaling exponents 
with respect to $< \dV^3 >$.
This distinction was not necessary in the bulk region where 
$\tau \equiv \hat \tau$.
By using expression (\ref{new_tau_p}) near the wall and eq. (\ref{zeta_p})
in the bulk region we obtain that the ``intermittency correction'' 
$\hat \tau(q)$ results to be essentially independent of the distance from the 
wall, fig. (\ref{F5.b}).
Hence the observed increase of intermittency of the velocity increments seems
to be associated more to the structure of the RKSH rather than to the 
intermittency of dissipation.
These theoretical findings seem to be confirmed by experimental results in a
flat plate boundary layer obtained recently by Ciliberto and coworkers 
(private communication).
  
We like here to emphasize that, to verify the new RKSH, we selected on purpose 
the plane closest to the wall where scaling laws still appear. 
On the opposite, in the bulk region, the original RKSH holds.
At intermediate planes we expect the scaling exponents to emerge 
from  a 
complex blending of these two basic behaviors, leading to a continuous 
variation with the distance from the wall \cite{Nizza}.

In conclusion, we have found that a quite evident failure of the RKSH occurs
in the near wall turbulence in correspondence with the simultaneous appearance 
of scaling laws. 
The new form of the RKSH we have proposed in this letter for the wall region 
is expressed in terms of the structure function of order two, 
instead of the structure function of order three as in the original form.  
This may be seen as a statistical representation of the physical features
of the near wall region, which is controlled more by the mechanism of
momentum transfer rather than by the classical energy cascade.

\acknowledgements
We acknowledge very useful discussions with L. Biferale, S. Succi and
S. Ciliberto.

%
%

\newpage
\newpage


\begin{thebibliography}{99}

\bibitem{FDR}
 {Amati, G\@., and Succi, S\@., and Piva, R\@.},
 {\it Preliminary analysis of the scaling exponents in channel flow turbulence},
 {Fluid Dyn. Res.}, {1998}.

\bibitem{Nizza}
  {Amati, G\@., and Succi, S\@., Piva, R\@., and Toschi, F\@.},
  {\it Scaling exponents in turbulent channel flow},
  {European Turbulence Conference VII}, {7}, {1998}, {Nizza}.

\bibitem{PRO}
 {Arad, I\@., Dhruva, B\@., Kurien, S\@., L'vov V\@. S\@., Procaccia, I\@. and
  Sreenivasan, K\@.R\@. },
 {\it The extraction of anisotropic contributions in turbulent flows}
 submitted to {Phys. Rev. Let.}, {1998}.

\bibitem{PhysD}
 {Benzi, R\@., Biferale, L\@., Ciliberto, S\@., Struglia, M\@. V\@. 
 and Tripiccione, R\@.},
 {\it Generalized scaling in fully developed turbulence},
 {Physica D}, {1996}, {96}, {162 -- 181}.

\bibitem{CA}
 {Casciola, C\@.M\@., and Amati, G\@.},
 {\it Helicity structures and intermittency in a channel flow},
 {8th Int. Symp. on Flow Visualization}, {1998}.


\bibitem{FRISH}
  {Frish, U\@.},
  {\it Turbulence: the legacy of A.N. Kolmogorov}, 
  {Cambridge University Press}, {Cambridge}, {1995}.

\bibitem{HINZE}
  {Hinze, J\@.O\@.},
  {\it Turbulence}, 
  {McGraw-Hill}, {New York}, {1959}.

\bibitem{KMM}
   {Kim, J\@., Moin, P\@. and Moser, R\@.},
   {\it Turbulence statistics in a fully developed channel flow at low 
   Reynolds number},
   {J. Fluid Mech.}, {1987}, {177}, {133 -- 166}.

 \bibitem{K62}
  {Kolmogorov, A\@.N\@.},
  {\it A refinement of previous hypothesis concerning the local structure
  of turbulence in a viscous incompressible fluid at high Reynolds number},
  {J. Fluid Mech.}, {1962}, {82 -- 85}, {6}.

 \bibitem{KRAIC}
  {Kraichnan, R\@.H\@.},
  {\it On Kolmogorov's inertial-range theories},
  {J. Fluid Mech.}, {1974}, {305 -- 330}, {62}.


\end{thebibliography}
\end{document}